\documentclass{ws-procs11x85}
\usepackage{ws-procs-thm}
\usepackage{amsfonts}
\usepackage{amsmath}
\usepackage{bbm}
\usepackage{graphicx}
\usepackage{xspace}
\usepackage{xcolor}
\usepackage{mathtools,soul}
\usepackage{multirow}
\usepackage{booktabs}
\usepackage{color}
\usepackage{longtable,threeparttablex}
\usepackage{ulem}
\usepackage{subcaption}
\usepackage{hyperref}
\usepackage{xr}
\usepackage{wrapfig}

\newcommand{\NV}{\mbox{$\mathop{\mathtt{Node2Vec}}\limits$}\xspace}
\newcommand{\DW}{\mbox{$\mathop{\mathtt{DeepWalk}}\limits$}\xspace}
\newcommand{\AD}{\mbox{$\mathop{\mathtt{AD}}\limits$}\xspace}
\newcommand{\method}{\mbox{$\mathop{\mathtt{MPI}}\limits$}\xspace}
\newcommand{\baseline}{\mbox{$\mathop{\mathtt{BSL}}\limits$}\xspace}
\newcommand{\km}{\mbox{$\mathop{\mathtt{KM}}\limits$}\xspace}
\newcommand{\hr}{\mbox{$\mathop{\mathtt{HR}}\limits$}\xspace}

\newcommand{\cox}{\mbox{$\mathop{\mathtt{COX}\texttt{-}\mathtt{INHs}}\limits$}\xspace}
\newcommand{\ace}{\mbox{$\mathop{\mathtt{ACE}\texttt{-}\mathtt{INHs}}\limits$}\xspace}
\newcommand{\ache}{\mbox{$\mathop{\mathtt{AChE}\texttt{-}\mathtt{INHs}}\limits$}\xspace}
\newcommand{\acr}{\mbox{$\mathop{\mathtt{AChR}\texttt{-}\mathtt{Ags}}\limits$}\xspace}
\newcommand{\bbb}{\mbox{$\mathop{\mathtt{BBB}}\limits$}\xspace}
\newcommand{\bbbx}{\mbox{$\mathop{\mathtt{BBBx}}\limits$}\xspace}
\newcommand{\ach}{\mbox{$\mathop{\mathtt{ACh}}\limits$}\xspace}

\newcommand{\coxs}{\mbox{$\mathop{\mathtt{COX}\texttt{-}\mathtt{INH}}\limits$}\xspace}
\newcommand{\aces}{\mbox{$\mathop{\mathtt{ACE}\texttt{-}\mathtt{INH}}\limits$}\xspace}
\newcommand{\aches}{\mbox{$\mathop{\mathtt{AChE}\texttt{-}\mathtt{INH}}\limits$}\xspace}
\newcommand{\acrs}{\mbox{$\mathop{\mathtt{AChR}\texttt{-}\mathtt{Ag}}\limits$}\xspace}
\newcommand{\etal}{\text{et al}.}

\begin{document}

\title{Modeling Path Importance for Effective Alzheimer’s Disease Drug Repurposing}

\author{Shunian Xiang$^{1*}$,
    Patrick J. Lawrence$^{1*}$,
    Bo Peng$^2$,
    ChienWei Chiang$^1$, PhD,
    Dokyoon Kim$^3$, PhD,
     Li Shen$^3$, PhD, and
    Xia Ning$^{1,2,4\dag}$, PhD}

\address{
    $^1$Biomedical Informatics Department, The Ohio State University, Columbus, OH 43210, USA\\
     $^2$Computer Science and Engineering Department, The Ohio State University, Columbus, OH 43210, USA\\
     $^3$Department of Biostatistics, Epidemiology, and Informatics, University of Pennsylvania,
     Philadelphia, PA 19104 USA\\
     $^4$Translational Data Analytics Institute, The Ohio State University, Columbus, OH 43210, USA\\
    $^*$Co-first author; authors contributed equally to this work\\
    $^\dag$E-mail: ning.104@osu.edu\\
}

\begin{abstract}
Recently, drug repurposing has emerged as an effective and resource-efficient paradigm for \AD drug discovery.
Among various methods for drug repurposing, network-based methods have shown promising results as they
are capable of leveraging complex networks that integrate multiple interaction types,
such as protein-protein interactions, to more effectively identify candidate drugs.
However, existing approaches typically assume paths of the same length in the network
have equal importance in identifying the therapeutic effect of drugs.
Other domains have found that same length paths do not necessarily have the same importance.
Thus, relying on this assumption may be deleterious to drug repurposing attempts.
In this work, we propose \method (Modeling Path Importance), a novel network-based method for \AD drug repurposing.
\method is unique in that it prioritizes important paths via learned node embeddings,
which can effectively capture a network's rich structural information.
Thus, leveraging learned embeddings allows \method to effectively differentiate the importance among paths.
We evaluate \method against a commonly used baseline method that identifies anti-\AD drug candidates primarily
based on the shortest paths between drugs and \AD in the network.
We observe that among the top-50 ranked drugs,
\method prioritizes 20.0\% more drugs with anti-\AD evidence compared to the baseline.
Finally, Cox proportional-hazard models produced from insurance claims data
aid us in identifying the use of etodolac, nicotine, and \bbb-crossing \ace as having a reduced risk of \AD,
suggesting such drugs may be viable candidates for repurposing and should be explored further in future studies.
\end{abstract}


\vspace{10pt}
\copyrightinfo{\copyright\ 2023 The Authors. Open Access chapter published by World Scientific Publishing Company and distributed under the terms of the Creative Commons Attribution Non-Commercial (CC BY-NC) 4.0 License.}
%

\section{Introduction}
\label{sec:introduction}

Alzheimer's Disease, denoted \AD, is a progressive neurodegenerative disorder that accounts for 60\%-70\% of dementia cases
and affects more than 50 million people worldwide today~\cite{silva2019alzheimer,passeri2022alzheimer}.
Given the large number of affected individuals and \AD's life-threatening nature~\cite{athar2021recent},
extensive resources have been dedicated to developing \AD-modifying drugs.
Since 2003, inefficacy or toxicity has accounted for a 95+\% failure rate
among candidates evaluated for \AD treatment~\cite{yu2021novel,kim2022alzheimer}.
Furthermore, none of the current US Food and Drug Administration (FDA)-approved \AD drugs are curative;
they only slow disease progression.
Because of the immense resources required to conduct clinical trials~\cite{cummings2022costs},
the numerous failed clinical trials have necessitated the development of a more
resource-efficient method for \AD drug discovery.
In the last decade, the identification of new therapeutic indications for existing FDA-approved drugs, referred to as drug repurposing~\cite{pushpakom2019drug},
has emerged as an effective and resource-efficient paradigm for drug discovery~\cite{zhan2022drug}.
This is an attractive option as the toxicity, pharmacokinetics, and pharmacodynamics of FDA-approved drugs have already been thoroughly investigated by previous clinical trials~\cite{pushpakom2019drug,begley2021drug}.

Recently, the curation of comprehensive drug databases has enabled the development of computational methods for \AD drug repurposing~\cite{park2019review,sosa2022contexts,morselli2021network,cheng2018network,cai2021drug}.
Among all the methods,
network-based methods have shown promising results and emerged as a popular approach~\cite{cheng2018network,savva2022network,fang2020network}.
%
Network-based methods utilize comprehensive  protein-protein, drug-target, and \AD-protein interactions to effectively reveal potential therapeutic effects of drugs on \AD.
%
Though promising, existing methods~\cite{cheng2018network} measure the therapeutic effects of drugs on \AD primarily using
count and length of the paths connecting drug nodes and the \AD node in the network.
Paths of the same length are considered equivalently effective at identifying the therapeutic effect of drugs by these methods.
However, in other domains, paths of the same length have been shown to exhibit substantially different levels of importance~\cite{sun2011pathsim,velickovic2018graph}.
As such, assuming equal length paths have equal importance could be detrimental to effective drug repurposing for \AD.

In this work, we propose a novel method to conduct drug repurposing for \AD, \method (Modeling Path Importance), to address this limitation.
Similar to existing methods~\cite{cheng2018network, cai2021drug}, \method leverages the interactions between drugs and \AD via proteins
as indications of the potential therapeutic effects of drugs on \AD.
Based on the interactions, \method introduces a scoring function to score and rank drugs for their anti-\AD effectiveness.
\method is unique in that it learns node embeddings~\cite{abu2018watch}
and prioritizes important paths via these learned embeddings.
Recent work~\cite{grover2016node2vec} has shown that the learned node embeddings can effectively capture the rich structure information within a network.
Thus, scoring paths using node embeddings allows \method
to utilize the network structure information to better
prioritize paths for effective \AD drug repurposing.
Specifically, in this study, \method leverages \DW~\cite{perozzi2014deepwalk},
 a widely used network learning approach, to generate node embeddings.
Edges are scored using a normalized dot product between the learned node embeddings;
paths and drugs are scored by multiplying individual edge scores.
Note that because \method serves as a general framework, other network learning approaches,
such as \NV~\cite{grover2016node2vec} and graph neural networks~\cite{kipf2016semi},
could also be easily incorporated to generate node embeddings.

In this study, we construct a network to conduct drug repurposing for \AD
by combining protein-protein interactions (PPIs), drug-target interactions (DTIs),
and \AD-protein interactions (APIs) from multiple data sources.
To investigate the effectiveness of \method,
we compare \method against a commonly utilized
network-based drug repurposing method for \AD~\cite{cheng2018network,fang2020harnessing},
denoted as \baseline, using our network.
Our experimental results demonstrate that among the top-50 ranked drugs,
\method prioritizes 20\% more drugs with anti-\AD evidence compared to \baseline.
We examine published literature and analyze insurance claims meta data
to evaluate the evidence of anti-\AD activity among \method's top prioritized candidates.
The results of our evaluation find consensus between published experimental results and our own analysis for a few drug candidates.
Notably, angiotensin converting enzyme inhibitors (\ace) represent a class of drugs
that should be further explored for their anti-AD properties.
Moreover, other drugs, such as nicotine, that enhance the brains response to acetylcholine
and reduce cholinergic atrophy should be examined as well.
Conversely, we find that, relative to other evaluated drugs,
long-term use of trihexyphenidyl increases the risk of \AD.
This was corroborated by previously published \textit{in vivo} experiments~\cite{huang2016long}.
Finally, we find etodolac to confer the lowest risk of developing \AD among all cyclooxygenase inhibitors (\cox) in our network.
Altogether, these findings suggest that \method may be a viable option with respect to
identifying repurposing candidates to treat \AD.

\section{Materials and Methods}
\label{sec:material}
\subsection{Network construction}
\label{sec:material:network}
PPIs, DTIs and APIs have shown utility for \AD drug repurposing~\cite{cheng2018network}.
As such, we construct our network using these interactions.
Below, we describe our process for compiling
the PPIs, DTIs and APIs used to construct our network from public data sources.
In total, our network has 327,924,  2,854, and 230 edges corresponding to PPIs, DTIs, and APIs.
These edges connect one \AD node, 18,527 protein nodes, and 386 drug nodes.

\subsubsection{Protein-protein interactions (PPIs)}
\label{sec:material:network:ppi}

Following Chen~\etal~\cite{cheng2018network}, we include a comprehensive list of human PPIs consisting of 327,924 interactions.
This list aggregates a total of 21 bioinformatics and systems biology databases with combinations of five types of experimental evidence.
We refer the audience of interest to Chen~\etal~\cite{cheng2018network} for a detailed description of the databases.
%
%
\subsubsection{Drug-target interactions (DTIs)}
\label{sec:material:network:dti}
We assemble drug-target interactions and
bioactivity data from 4 commonly used databases (each downloaded in November 2022):
the ChEMBL database~\cite{mendez2019chembl} (v31),
the binding database~\cite{liu2007bindingdb},
the therapeutic target database~\cite{chen2002ttd},
and the IUPHAR/BPS guide to pharmacology database~\cite{harding2022iuphar}.
We retain the drug-target interactions that satisfy
the following inclusion criteria:
1) binding affinities, including
$\text{K}_i$, $\text{K}_d$, $\text{IC}_{50}$, or $\text{EC}_{50}$, must be less than or equal to 10 $\mu$M;
2) protein targets and their respective proteins must have a unique UniProt~\cite{apweiler2004uniprot} accession number;
3) protein targets must be marked as reviewed in the UniProt database;
4) protein targets must be present in \textit{homo sapiens}.

Additionally, we retain drugs for which we have sufficient sample size
to conduct quantitative analysis using MarketScan~\cite{marketscan} insurance claims meta data
(see Section~\ref{sec:material:cox}).
Specifically, included drugs have at least 100 patients with their
first dose at least 2 years prior to an \AD diagnosis (dx).
Additionally, these drugs must have at least 15 patients who eventually received an \AD dx.
Applying these filters yielded
2,854 edges connecting 386 FDA-approved drugs to 548 protein targets.

\subsubsection{\AD-protein interactions (APIs)}
\label{sec:material:network:agi}

The \AD-associated proteins included in the network
were identified from multiple sources.
\mbox{54 $\beta$-amyloid-related} proteins and 27 tauopathy-related proteins were obtained from
Cheng \etal~\cite{cheng2018network}.
The authors identified proteins
that satisfied at least one of the following criteria:
1) the proteins are validated
in large-scale amyloid or tauopathy genome-wide association studies;
2) \textit{in vivo} experimental models exhibit evidence that knockdown or overexpression of
the protein leads to \AD-like amyloid or tau pathology.
We also include 93 unique
late-onset \AD common risk proteins
identified by 7 large-scale genetic studies
~\cite{lambert2013meta,marioni2018gwas,jansen2019genome,kunkle2019genetic,de2021common,wightman2021genome,bellenguez2022new}.
We further incorporate a set of 118 \AD-associated proteins introduced in at least 2 out of the 6 following databases (each was downloaded in November 2022):
the online Mendelian inheritance in man database~\cite{hamosh2005online},
the comparative toxicogenomics database~\cite{davis2009comparative},
the HuGE navigator database~\cite{yu2008navigator},
the DisGeNET database~\cite{pinero2016disgenet} (v7.0),
the ClinVar database~\cite{landrum2018clinvar}
and the Open Targets database~\cite{ochoa2023next} (v22.09).
In total, our network is comprised of 230 unique,
\AD-associated proteins.
Each of the \AD-associated proteins are
connected to a single \AD node with each edge between a protein and the \AD node representing an API in our network.

\subsection{Modeling path importance for \AD drug repurposing}
\label{sec:material:method}
In this work,
we denote the constructed network as $G$.
Each node in $G$ is denoted as $v_i$.
Specifically, drug nodes, protein nodes and the \AD node are
$v^d_i$, $v^g_i$, and $v^a_i$
, respectively.
Note that the index, $i$, does not apply to the \AD node as there are not multiple in our network.
Each edge that connects node $v_i$ to node $v_j$
is denoted as $e_{ij}$.
%
Each path is denoted as $p_m$, and the set of edges involved in a path is denoted $\mathbb{E}_{p_m}$.
Below, we denote matrices, scalars and row vectors using uppercase,
lowercase, and bold lowercase letters, respectively.
In \method, we leveraged \DW~\cite{perozzi2014deepwalk},
a widely used node embedding approach,
to learn embeddings for each node in $G$.
First, for each node $v_i$ in the network,
we conduct 256 random walks originating from this node,
and terminating once the path length reaches 128.
\DW is then trained by sliding a window of length 10 over the generated paths.
Nodes within the same window are forced to have similar embeddings following the objective function defined in the original paper~\cite{perozzi2014deepwalk}.
Node embeddings for \method are produced such that they have 128 dimensions.

After generating node embeddings, we score edge, $e_{ij}$,
using a normalized dot product of the embedding of
$v_i^x$ ($x$=$d$, $g$ or $a$) and $v_j^y$ ($y$=$d$, $g$ or $a$) as follows:
\begin{equation}
    \label{eqn:softmax}
    w_{ij} = \frac{\exp(\mathbf{v}_i^x {\mathbf{v}_j^y}^\top)}
    {\sum_{k \in \mathbb{V}}
    \exp(\mathbf{v}_i^x {\mathbf{v}_k^y}^\top)},
\end{equation}
where $w_{ij}$ is the score of the edge $e_{ij}$;
$\mathbf{v}_i^x$ and $\mathbf{v}_j^y$
is the learned embedding of node $v_i^x$ and $v_j^y$, respectively;
$\exp(\cdot)$ is the exponential function;
and $\mathbb{V}$ is the set of all the nodes in the network.
Note that, in Equation~\ref{eqn:softmax}, only one of $v_i^x$ and $v_j^y$ could be the \AD node.
These edge scores are calculated with node embeddings which implicitly capture the rich structural information within the network.
Thus, compared to existing methods, \method can better leverage a network's structural information for \AD drug repurposing.
%
%
We calculate the score for each path by multiplying the scores of its individual edges as follows:
\begin{equation}
    \label{eqn:path}
    s_{p_m} = \prod_{e_{ij} \in \mathbb{E}_{p_m}} w_{ij},
\end{equation}
where $s_{p_m}$ is the score of the path $p_m$;
and $\mathbb{E}_p$ is the set of all the edges in the path $p$.
The score for each drug (i.e., $s_{v_i^d}$) is then defined as
the summation of the scores from all 3-hop or shorter paths
that originate from the \AD node and terminate at the drug node.
%

\subsection{Baseline method}
\label{sec:material:baseline}

To evaluate the performance of \method,
we compare \method against a network-based method recently developed by Cheng \etal~\cite{cheng2018network},
denoted as \baseline.
\baseline scores drugs based on the shortest distance between the drug targets and the \AD-associated proteins (Section~\ref{sec:material:network:agi}) in the network.
Specifically, we denote $\mathbb{T}(i)$ as the set of protein targets associated with a given drug $v_i^d$, and denote $\mathbb{P}$ as the set of \AD-associated proteins.
The proximity between these two sets is calculated as the average shortest distance between elements
in $\mathbb{T}(i)$ and $\mathbb{P}$ as follows:
\begin{equation}
    \label{eqn:baseline}
    r(\mathbb{T}(i), \mathbb{P}) =
    \frac{1}{|\mathbb{T}(i)| + |\mathbb{P}|}
    \left(\sum_{v_j \in \mathbb{T}(i)} \min_{v_k \in \mathbb{P}} d (v_j, v_k)
    + \sum_{v_k \in \mathbb{P}} \min_{v_j \in \mathbb{T}(i)} d (v_k, v_j)
    \right),
\end{equation}
where $r(\mathbb{T}(i), \mathbb{P})$ is the proximity between these two sets;
$|\mathbb{T}(i)|$ and $|\mathbb{P}|$ is
the size of $\mathbb{T}(i)$ and $\mathbb{P}$, respectively;
and $\min_{v_k \in \mathbb{P}} d (v_j, v_k)$ is the shortest distance between $v_j$ and any elements in $\mathbb{P}$.
Subsequently, we conduct a permutation test to assess the statistical significance of the calculated proximity.
The resulting z-score from this test is used as the score of drug $v_i$~\cite{cheng2018network}.
In \baseline, a lower drug score implies a higher potential for effective \AD treatment.
%
%
\subsection{Validation using MarketScan database}
\label{sec:material:cox}
We use MarketScan medicare supplemental database from 2012\---2021
to evaluate drug impact on \AD onset via Cox proportional-hazard models~\cite{marketscan}.
The MarketScan database includes data for over 8 million unique individuals
and is comprised of demographic information, administrative information, diagnoses, procedures, and pharmacy records.
International Classification of Disease (ICD)-9/ICD-10 codes denote diagnoses
and National Drug Codes (NDCs) record pharmacy claims.
We use the ICD-9/ICD-10 codes listed in Supplementary Table S4$^\spadesuit$ to define \AD and comorbidities,
which are included as covariates in Cox proportional-hazard models.
We conduct our analysis over 1,632,218 unique individuals who were at least 65 years by 2022
and possessed a minimum of five years insurance enrollment prior of first \AD diagnosis.
Drugs from our constructed network are mapped to NDC codes
by partial matching of generic names from MartketScan redbook.
We only include patients who took or started taking a drug at least two years prior to \AD diagnosis
to mitigate the possibility that patients starting a drug already had \AD given that \AD is difficult to diagnosis.

\copyrightinfo{$^{\spadesuit}$ Supplementary materials and code can be found here: \url{https://github.com/ninglab/MPI}}

\section{Results}
\label{sec:results}

\subsection{\method for \AD drug repurposing}
\label{sec:results:method}

\begin{figure}
    \centering
    \includegraphics[width=0.8\linewidth]{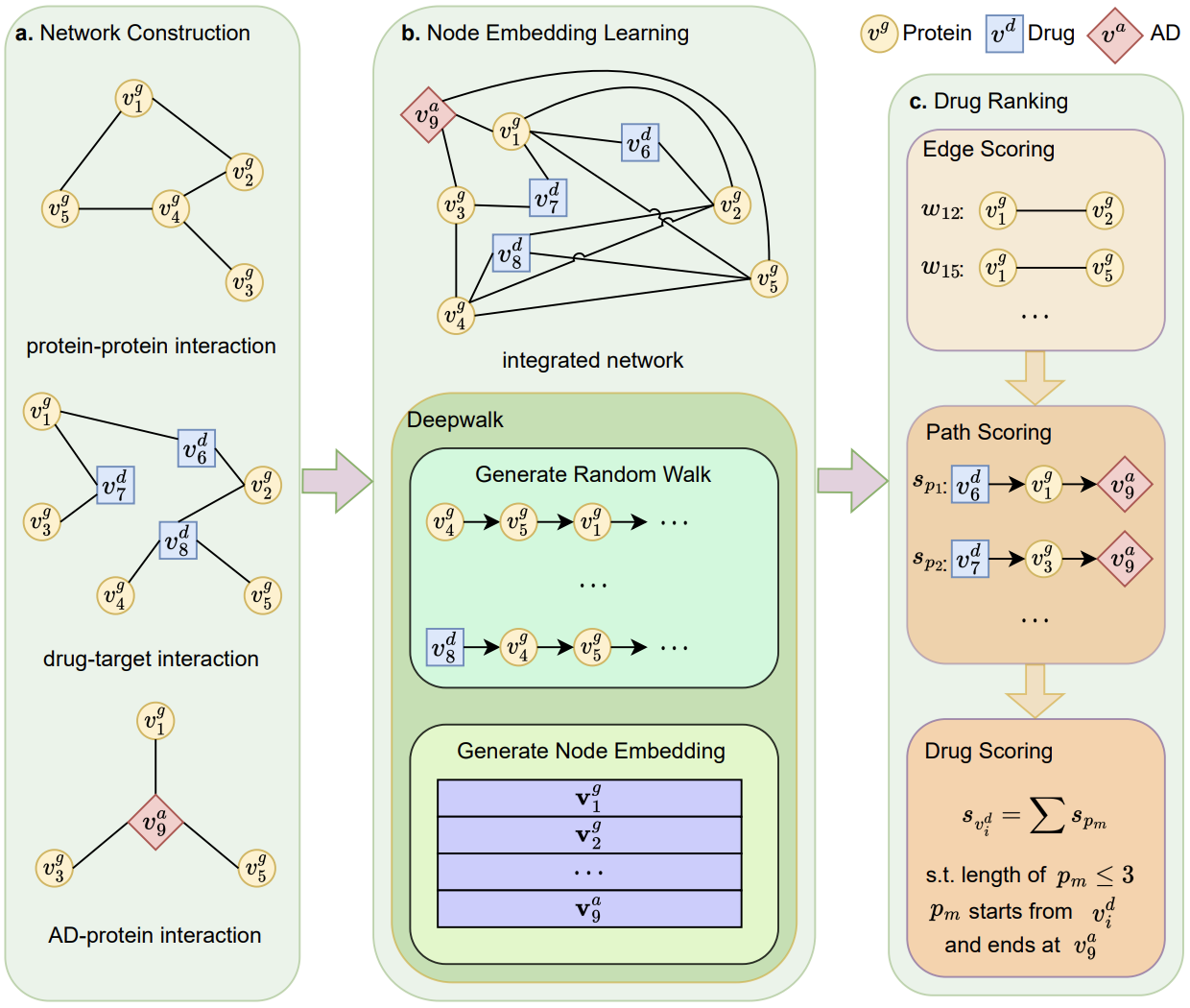}
    \caption{
    Figure~\ref{fig:overall}a shows the network construction process in \method.
    Figure~\ref{fig:overall}b shows the \DW-based node embedding generation in \method.
    Figure~\ref{fig:overall}c shows the edge, path and drug scoring in \method.
    }
    \label{fig:overall}
\end{figure}

In this study,
we curate a network consisting PPIs, DTIs and AGIs and
propose a novel network-based method, \method, for \AD drug repurposing.
We propose \method with the following intuitions:
1) proteins that associated with \AD are localized in the corresponding disease module within
the comprehensive human PPI network;
2) the drug target(s) for a disease may also be targeted for other diseases (e.g., \AD)
owing to common functional targets
and pathways elucidated by PPIs;
3) if a drug node is linked to the \AD node through the paths of drug targets and \AD-associated proteins in the PPI, the drug may have a treatment effect on \AD.
%

We implement \method using the following steps:
1) integrate \AD-protein interactions, drug-target interactions and protein-protein interactions
to generate a comprehensive network (Figure~\ref{fig:overall}a),
2) employ \DW to learn node embeddings which capture the structural information within the network (Figure~\ref{fig:overall}b), and
3) score edges, paths and drugs based on the learned embeddings to leverage the structural information for better \AD drug repurposing (Figure~\ref{fig:overall}c).
Then we identify plausible treatment candidates
from the top-ranked drugs using a literature search of the published evidence.
We collected 327,924 PPIs from 21 bioinformatics and systems biology databases (Section~\ref{sec:material:network:ppi}).
We also collected 2,854 DTIs
from 4 commonly used databases (Section~\ref{sec:material:network:dti}),
and 230 comprehensive APIs from multiple resources (Section~\ref{sec:material:network:agi}).
By aggregating all the interactions,
we construct a drug-protein-\AD network comprised of
386 drug nodes, 18,527 protein nodes, 1 \AD node, and 331,008 edges.
More details about the network construction are available in Section~\ref{sec:material}.
To the best of our knowledge,
\method is the first method which effectively repurposes drug candidates for \AD treatment
by prioritizing paths between drug nodes and the \AD node using learned node embeddings.
%


\begin{wrapfigure}{r}{0.5\linewidth}
    \begin{subfigure}{\linewidth}
        \centering
        \includegraphics[width=\linewidth]{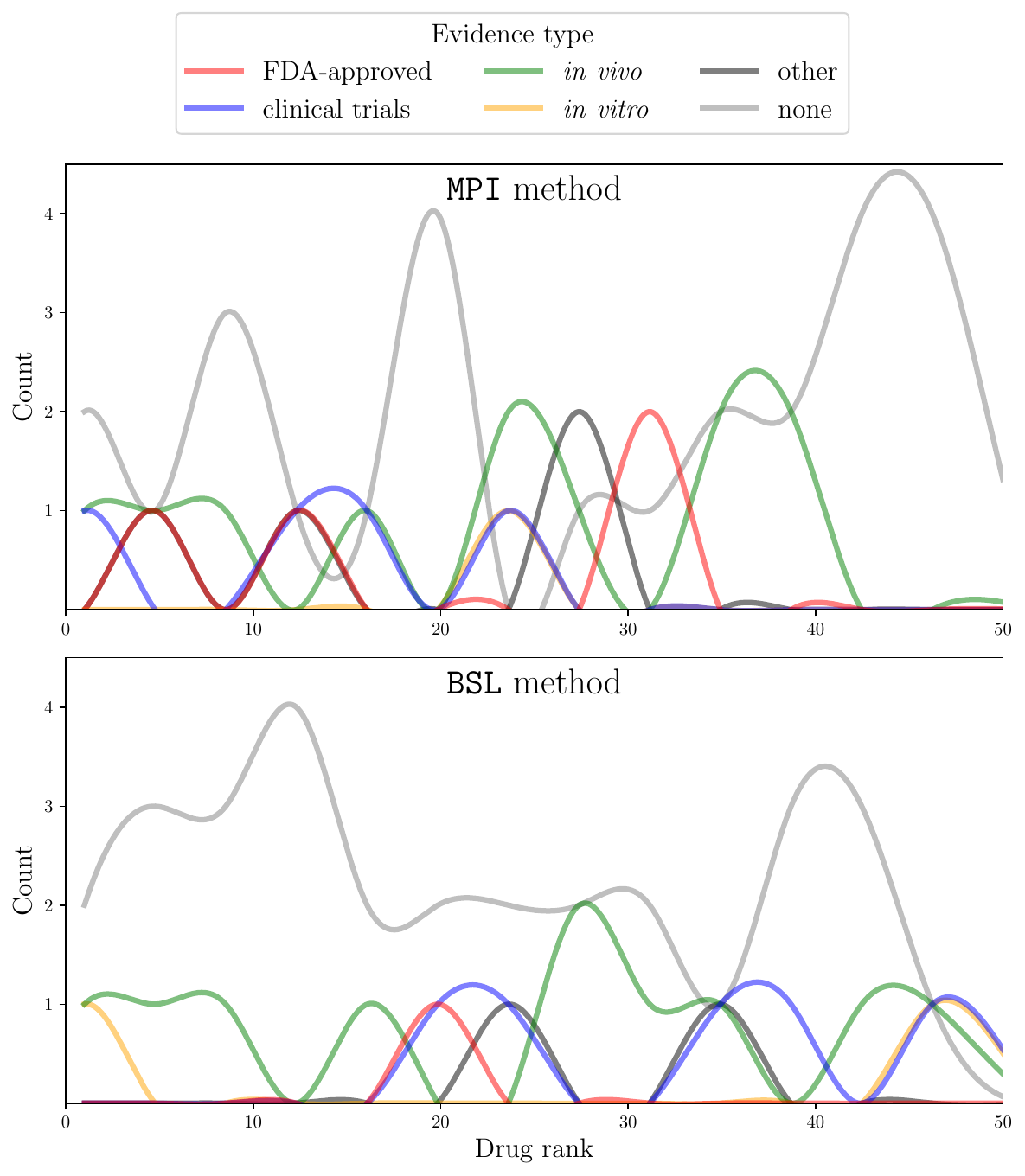}
        \caption{Distribution of drug rank by evidence type for \baseline and \method methods.}
        \label{fig:ranking}
    \end{subfigure}
\\
    \begin{subfigure}{\linewidth}
        \centering
        \includegraphics[width=\linewidth]{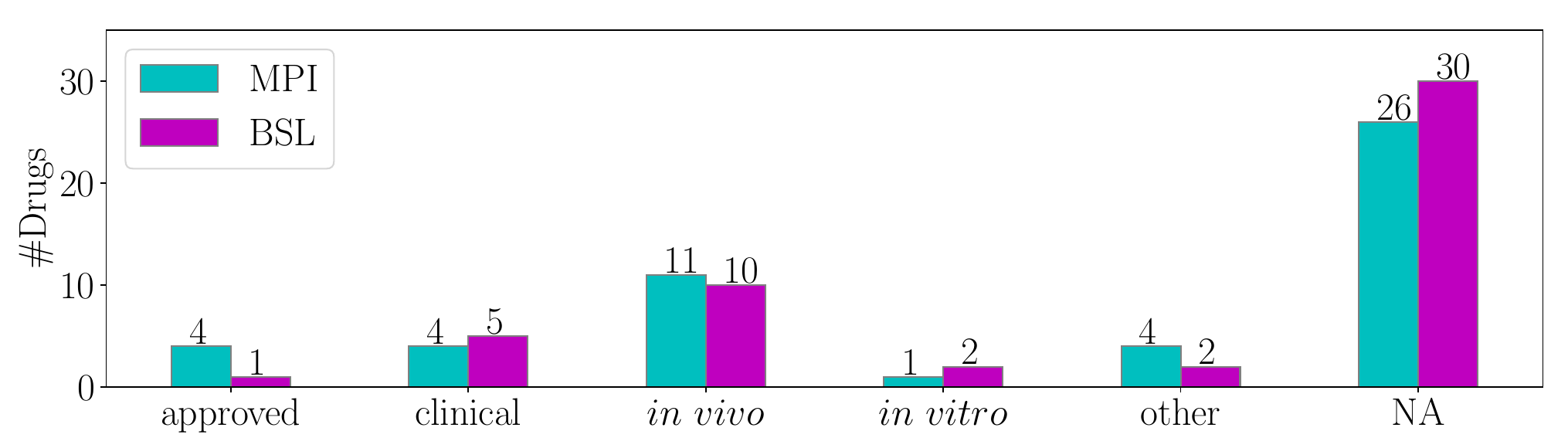}
        \caption{Drug counts for each type of evidence among \method's and \baseline's top-50 drugs.}
        \label{fig:evidence}
    \end{subfigure}
\caption{Evaluation of drug rank distributions: \method and \baseline.}
\end{wrapfigure}
\subsection{Comparing anti-\AD evidence of \method's and \baseline's top-50 drugs}
\label{sec:results:result}
We compare the top-50 drugs prioritized by \method and \baseline to evaluate their capacity for repurposing drugs to treat \AD.
Specifically, we score and rank all 386 drug nodes in our network using \method and \baseline.
The complete rankings are reported in Supplementary Table S3.
We then perform a literature search to evaluate the anti-\AD evidence of the top-50 ranked drugs for both \method and \baseline.
We define anti-\AD evidence as any published experimental result(s),
which demonstrate a drug either protects against the development of \AD
or ameliorates aberrant cellular phenotypes caused by \AD.
We present \method's and \baseline's top-10 drugs and their anti-\AD evidence
     in Table~\ref{tbl:drugs_method_10} and Table~\ref{tbl:drugs_baseline_10}, respectively.
The complete rankings for the top-50 drugs and their anti-\AD evidence is available in Supplementary Tables S1 and S2.
%
Based on the significance of the anti-\AD evidence,
we categorized drugs into the following 6 types
in decreasing order of significance:
1) drugs which are FDA-approved for \AD treatment (approved);
2) drugs that have demonstrated anti-\AD effects in completed clinical trials or are under investigation in \AD clinical trials (clinical);
3) drugs which have demonstrated anti-\AD effects in \mbox{\textit{in vivo}} experiments (\textit{in vivo});
4) drugs which have demonstrated anti-\AD effects in \textit{in vitro} experiments (\textit{in vitro});
5) drugs which show anti-\AD effects in observational studies, cohort studies or analyses in insurance data (other);
6) drugs that either do not have the above 5 types of evidence or have been demonstrated ineffective or damaging for \AD (NA).
We present the distribution of the top-50 drugs from \method and \baseline over the different types of evidence
in Figure~\ref{fig:ranking} and the counts of each evidence type in Figure~\ref{fig:evidence}.
In Figure~\ref{fig:ranking}, we observe more drugs with evidence ranked highly by \method compared to \baseline.
%
This is supported by Figure~\ref{fig:evidence}, which confirms that
\method identified more evidential anti-\AD drugs compared to \baseline in the top-50 ranked drugs.
Specifically, among the top-50 ranked drugs, \method prioritized 24 evidential anti-\AD drugs while \baseline only prioritized 20 evidential anti-\AD drugs, demonstrating an improvement of 20\%.
Figures~\ref{fig:ranking} and \ref{fig:evidence} also show \method outperforms \baseline in prioritizing drugs with significant evidence.
\method prioritizes all the 4 FDA-approved anti-\AD drugs (e.g., galantamine, rivastigmine, donepezil and memantine) in our network among the top-50.
In contrast, \baseline prioritizes only a single FDA-approved anti-\AD drug
(donepezil) among the top-50.

We also observe in Table~\ref{tbl:drugs_method_10} and Table~\ref{tbl:drugs_baseline_10} that
\method is more effective than \baseline at prioritizing anti-\AD drugs among the very top (top-10) of the ranking list.
That is, among the top-10 drugs, 6 drugs from \method have anti-\AD evidence including the FDA-approved \AD drug galantamine, while only 4 drugs from \baseline are evidential.
As presented in Section~\ref{sec:material}, compared to \baseline, \method learns node embeddings to capture the rich structural information within the network, and leverage the structural information to better identify anti-\AD drugs.
The superior performance of \method over \baseline demonstrates the effectiveness of leveraging the network structural
information to conduct repurposing to identify candidates for \AD treatment.
We also notice that both \method and \baseline prioritize 17 drugs in concordance within their top-50 drug lists.
Among the 17 drugs, 5 drugs demonstrate anti-\AD evidence:
donepezil is an FDA-approved anti-\AD drug;
nicotine and rasagiline have  clinical anti-\AD evidence;
and fluvoxamine and fluoxetine have \mbox{\textit{in vivo}} anti-\AD evidence.
The drugs nicotine, rasagiline, fluvoxamine, and fluoxetine could be promising repurposing candidates. We leave the investigation of these drugs to future research.

\vspace{5pt}


\begin{table}
\footnotesize
  \caption{Top-10 Drugs from \method}
  \centering
  \label{tbl:drugs_method_10}
  \begin{threeparttable}
      \begin{tabular}{
	@{\hspace{8pt}}l@{\hspace{8pt}}
	@{\hspace{8pt}}l@{\hspace{8pt}}
	@{\hspace{8pt}}l@{\hspace{8pt}}
	@{\hspace{8pt}}c@{\hspace{8pt}}
	@{\hspace{8pt}}l@{\hspace{8pt}}
	}
    \toprule
    Drug & MOA & Indication & Anti-\AD & Evidence\\
    \midrule
    varenicline & \acrs & smoking cessation & N & -\\
    fosinopril &
    \aces & hypertension & Y & \textit{in vivo}~\cite{deb2015comparative}\\
    nicotine & \acrs
    & smoking cessation & Y & clinical~\cite{newhouse2012nicotine}\\
    nizatidine & histamine receptor antagonist & duodenal ulcer disease & N & -\\
    piroxicam & \coxs & osteoarthritis & Y & other~\cite{zhang2018nsaid,imbimbo2010nsaids}\\
    meloxicam & \coxs & osteoarthritis & Y & \textit{in vivo}~\cite{ianiski2016meloxicam,guan2023meloxicam,ianiski2012protective}\\
    galantamine & \aches & Alzheimer's disease & Y & approved\\
    bromfenac & \coxs & inflammation & N & -\\
    etodolac & \coxs & osteoarthritis & Y & \textit{in vivo}~\cite{elfakhri2019multi}\\
    pyridostigmine & \aches & myasthenia gravis & N & -\\
    \bottomrule
    \end{tabular}
    \begin{tablenotes}[normal,flushleft]
    \item{\footnotesize{In this table, the column ``Drug" shows the identified top-10 ranked drugs; the column ``MOA" shows the mechanism of action of each drug; the column ``Indication" presents the indication of each drug; the column ``Anti-\AD" indicates if the drug has evidenced anti-\AD effects;
    and the column ``Evidence" presents the type of the evidence.
    In this table, \aces represents the angiotensin converting enzyme inhibitor;
    \coxs represents the cyclooxygenase inhibitor;
    \aches represents the acetylcholinesterase inhibitor;
    and \acrs represents the acetylcholine receptor agonist.}}
    \par
    \end{tablenotes}
  \end{threeparttable}
  \vspace{-10pt}
\end{table}


\begin{table}
\footnotesize
  \caption{Top-10 Drugs from \baseline}
  \centering
  \label{tbl:drugs_baseline_10}
  \begin{threeparttable}
      \begin{tabular}{
	@{\hspace{8pt}}l@{\hspace{8pt}}
	@{\hspace{8pt}}l@{\hspace{8pt}}
	@{\hspace{8pt}}l@{\hspace{8pt}}
	@{\hspace{8pt}}c@{\hspace{8pt}}
	@{\hspace{8pt}}l@{\hspace{8pt}}
	}
    \toprule
    Drug & MOA & Indication & Anti-\AD & Evidence\\
    \midrule
    \multirow{2}{*}{tetracycline} & bacterial 30S
    & respiratory tract & \multirow{2}{*}{Y}
    & \multirow{2}{*}{\textit{in vitro}~\cite{forloni2001anti}}\\
    & ribosomal subunit inhibitor & infections &&\\
    selegiline & monoamine oxidase inhibitor & Parkinson's Disease & N & -\\
  \multirow{2}{*}{ceftriaxone} & bacterial cell wall & \multirow{2}{*}{gonorrhea} & \multirow{2}{*}{Y} & \multirow{2}{*}{\textit{in vivo}~\cite{tikhonova2021neuroprotective}}\\
  & synthesis inhibitor &&&\\
  ibuprofen & \coxs & headache & N & -\\
  levobunolol & adrenergic receptor antagonist & glaucoma & N & -\\
  ketoprofen & \coxs & rheumatoid arthritis & N & -\\
  \multirow{2}{*}{carbidopa} & aromatic L-amino acid & \multirow{2}{*}{Parkinson's Disease} & \multirow{2}{*}{N} & \multirow{2}{*}{-}\\
  & decarboxylase inhibitor &&&\\
  sulindac & \coxs & osteoarthritis & Y & \textit{in vivo}~\cite{modi2014sulindac}\\
  biotin & vitamin B & supplement & Y & \textit{in vivo}~\cite{lohr2020biotin}\\
  lansoprazole & ATPase inhibitor & heartburn & N & -\\
    \bottomrule
    \end{tabular}
    \begin{tablenotes}[normal,flushleft]
    \item{\footnotesize{In this table, the column ``Drug" shows the identified top-10 ranked drugs; the column ``MOA" shows the mechanism of action of each drug; the column ``Indication" presents the indication of each drug; the column ``Anti-\AD" indicates if the drug has evidenced anti-\AD effects;
    and the column ``Evidence" presents the type of the evidence.
    In this table, \aces represents the angiotensin converting enzyme inhibitor;
    \coxs represents the cyclooxygenase inhibitor;
    \aches represents the acetylcholinesterase inhibitor;
    and \acrs represents the acetylcholine receptor agonist.}}
    \par
    \end{tablenotes}
  \end{threeparttable}
  \vspace{-10pt}
\end{table}


\subsection{Identifying repurposing candidates with anti-\AD activity}
\label{sec:results:drug}
\begin{figure}[!h]
    \centering
    \footnotesize
    \begin{minipage}{\linewidth}
        \begin{subfigure}{0.32\linewidth}
            \centering
            \includegraphics[width=\linewidth]{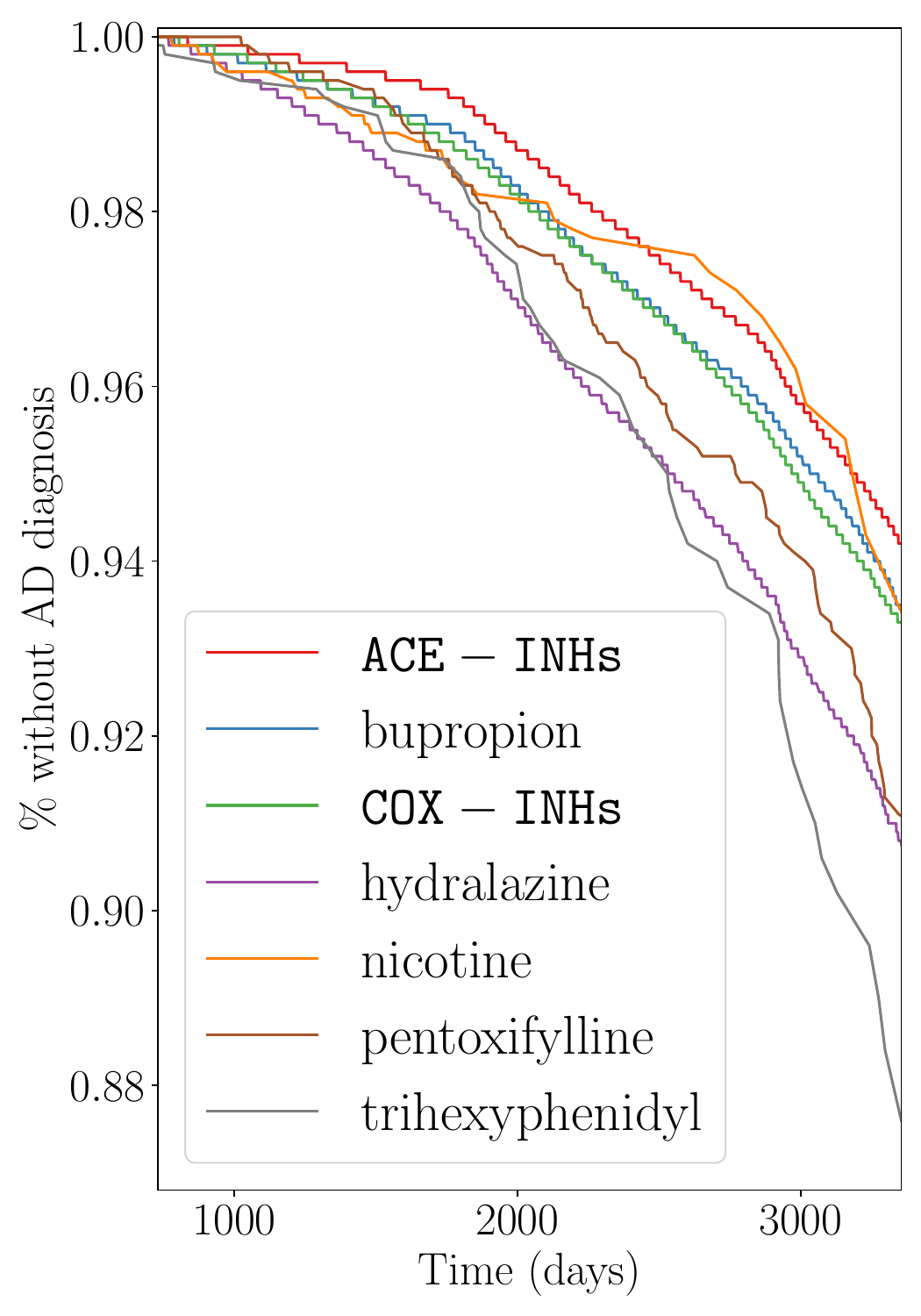}
            \caption{Drugs and MOAs with \\published anti-\AD evidence\\}
            \label{fig:cox}
        \end{subfigure}
        \begin{subfigure}{0.32\linewidth}
            \centering
            \includegraphics[width=\linewidth]{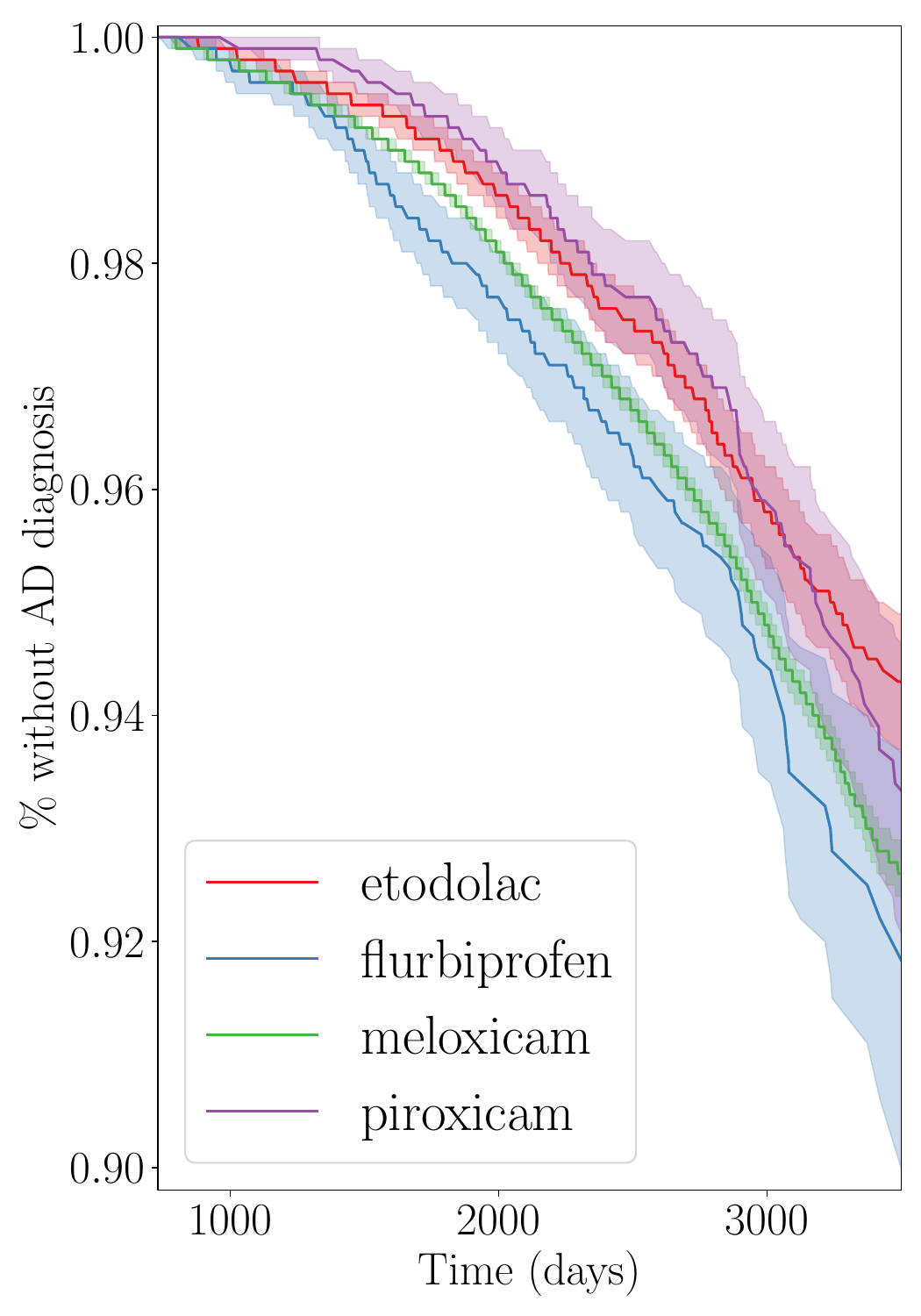}
            \caption{\cox. Shaded regions represent 95\% confidence\\intervals.}
            \label{fig:coxinh}
        \end{subfigure}
        \begin{subfigure}{0.32\linewidth}
            \centering
            \includegraphics[width=\linewidth]{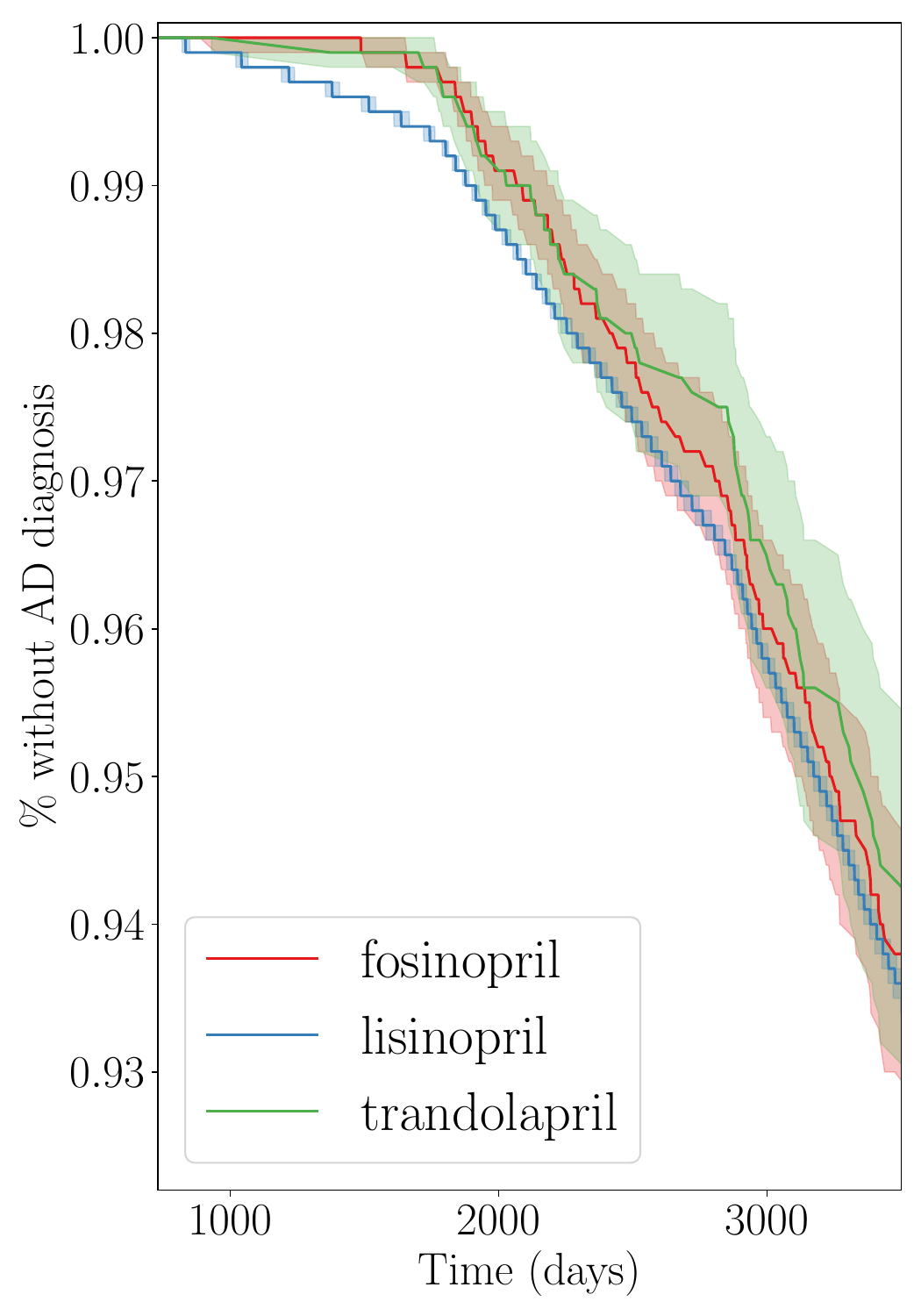}
            \caption{\ace. Shaded regions represent 95\% confidence\\intervals.}
            \label{fig:aceinh}
        \end{subfigure}
    \end{minipage}
    \caption{Unadjusted Kaplan$\--$Meier plots for cox proportional-hazard models}
    \label{fig:domain}
    \vspace{-15pt}
\end{figure}

%
In order to identify plausible candidates for repurposing,
we produce Cox proportional-hazard models (see Section~\ref{sec:material:cox})
to ascertain whether there is consensus between the MarketScan insurance data
and the \AD-related evidence we found for top ranked candidates prioritized by \method.
Specifically, we use hazard ratios (\hr) to identify whether any evidential drug elicited reduced the risk of \AD diagnosis among patients who took the drug compared against those that did not.
We present each drug's \hr with their significance levels in {Supplementary Table S5b};
the \hr for each drug's covariates (sex, age, and additional common comorbidities)
are reported in Supplementary Table S5c-t.
A {\hr} below 1 indicates that a drug has a protective effect,
while a {\hr} above 1 indicates that a drug has a damaging effect.
Figure~\ref{fig:cox} plots {Kaplan-Meier} (\km) survival curves.
These plots depict a patient's likelihood of being diagnosed with \AD following long-term use of
either an individual prescribed drug or a drug with a given mechanism of action (MOA).
For MOAs, we group highly-prioritized drugs with
published evidence of anti-\AD activity (see Table~\ref{tbl:drugs_method_10}).
Bupropion ($\hr=1.04$; non-significant) was included as a negative control
as clinical trials found the drug had no significant effect on cognition in AD patients~\cite{maier2020bupropion}.
 Trihexyphenydil ($\hr=1.71;\ \alpha<0.001$) was included as a positive control for damaging effects
 due to the evidence documented in Supplementary Table S1.
The \cox group includes the following drugs:
piroxicam, meloxicam, etodolac, and flurbiprofen.
The \ace group includes the following drugs:
fosinopril, trandolapril, and lisinopril.
Note that we only include blood brain barrier (\bbb) crossing \ace in this group
as non-\bbb-crossing \ace have exhibited very limited effects on \AD~\cite{ouk2021use}.
We also include {time-to-event} analysis for 4 of \baseline's top prioritized drugs (See Supplementary Figure S1).
Unlike \method, we observe only one of \baseline's drugs (sulindac) with reduced {time-to-event} compared to bupropion;
however, this difference is not significant.
\vspace{-10pt}
%

\subsection{Analyzing the MOAs of \method's top-50 drugs}
\label{sec:results:moa}

To identify groups of drugs whose anti-\AD properties should be further examined and explored,
we examine the top-50 drugs prioritized by \method for any common MOAs.
We find that \cox and \ace are the most common MOAs prioritized by \method.
Both \cox and \ace have published evidence of anti-\AD activity.
That said, experimental results suggest that long-term administration of \cox
may only have protective properties, reducing the risk of \AD onset~\cite{zhang2018nsaid}.
Moreover, meloxicam ($\hr=0.86;\ \alpha<0.05$), has even shown therapeutic potential, reversing
cognitive decline via inhibition of neuronal apoptosis~\cite{guan2023meloxicam,ianiski2016meloxicam}.
However, in Figure~\ref{fig:cox},
we observe that \cox as a class do not yield reduced risk of \AD
compared to the negative control.
That said, we find etodolac significantly reduces
the risk of \AD ($\hr=0.78;\ \alpha<0.001$) compared to other \cox,
including flurbiprofen ($\hr=0.95$; non-significant) (Figure~\ref{fig:coxinh}).
This suggests that only certain \cox, such as etodolac, may elicit protective effects against \AD onset.
Importantly, this may be a result of differences in target as etodolac targets \textit{COX2},
while flurbiprofen targets \textit{COX1}.
On the other hand, \ace were found to also protect against \AD onset in Figure~\ref{fig:cox}.
Specifically, we evaluate only \ace that cross the blood brain barrier (\bbb) as previous insurance claims metadata analyses have indicated those that do not cross the \bbb have no effect on \AD~\cite{ouk2021use}.
To see if any of the \bbb crossing \ace have a greater protective effect that others,
we produce a \km plot for fosinopril, lisinopril, and trandolapril (Figure~\ref{fig:aceinh}).
Unlike for \cox, \ace do not elicit any significant by-drug difference in \AD onset
as illustrated in Figure~\ref{fig:aceinh}.
While \method prioritized four \bbb crossing (\bbbx) and
four non-\bbbx \ace in the top-50,
the \bbbx \ace had a lower average rank compared to the non-\bbbx \ace (15 and 19, respectively).

Another important distinction between \cox and \ace is that \ace have been shown to have some ameliorative potential;
whereas, \cox have only shown protective effects.
In fact, fosninpril and lisinopril (ranked $2^{nd}$ and $24^{th}$ by \method, respectively) was found to reduce cognitive decline in animal models of \AD~\cite{deb2015comparative,thomas2021angiotensin}.
In Figure~\ref{fig:cox}, we find that \bbbx \ace consistently exhibit decreased risk of \AD
relative to our negative control drug, bupropion.
Additionally, there does not appear to be a significant difference between any of the \bbbx \ace
with respect to their protection against \AD, indicating that they are possibly all viable candidates for repurposing.
This is in agreement with other published evidence that has identified \bbbx \ace
as having protective effect on \AD development.
Interestingly, \method prioritized 133.6\% more \cox and 700.9\% more \ace than \baseline
in the top-50 from all such drugs in our network.
\method's ability to prioritize more drugs from MOAs with known anti-AD activity
suggests that it may be a more viable option when identifying candidates for drug repurposing.

\method also highly prioritizes drugs that increase the brain's response to acetylcholine, either by reducing its degradation (acetylcholinesterase inhibitors, \ache) or by stimulating its receptors (acetylcholine receptor agonists, \acr).
This is important as acetylcholine's (\ach) synaptic bioavailability is an important contributor to \AD progression.
That is, there is evidence that cholenergic atrophy and \ach deficiency is linked with cognitive decline in \AD patients~\cite{knowles1982denitrification}.
Moveover, many of the current FDA-approved drugs indicated to slow \AD progression target this mechanism of disease progression via \ache (e.g., donepezil, rivastigmine, and galantamine).
\acr, also enhances \ach signaling.
Such drugs, such as nicotine,
accomplish this by increasing the response of \ach receptors located on the post-synaptic neuron.
Interestingly, nicotine, was found to significantly improve cognition
in patients with mild cognitive impairment, which is a precursor to \AD~\cite{newhouse2012nicotine}.
We also find long-term nicotine use to have a protective effect ($\hr=0.532;\ \alpha<0.001$),
 with respect to \AD onset.
In Figure~\ref{fig:cox}, we observe similar risk of developing \AD to \ace after six to seven years.
Conversely, we find evidence that long-term use of trihexyphenidyl,
which reduces the activity of \ach receptors,
is associated with \AD-like neurodegeneration in rats~\cite{huang2016long}.
This is corroborated by Figure~\ref{fig:cox},
where we observe the highest risk of \AD
elicited by trihexyphenidyl.
More than eight years on trihexyphenidyl was associated
with a substantial increase in the risk of \AD
relative to the other drugs evaluated in Figure~\ref{fig:cox}.
These findings confirm that \ach signaling is closely linked with \AD progression.
As such, exploring other drugs and drug classes which either
increase \ach synaptic bioavailability or enhance neuronal response to \ach
should be further examined for anti-\AD activity.
%
%

\section{Discussion}
\label{sec:discussion}
In this work, we propose a novel network-based, \AD-specific drug repurposing approach called \method.
\method improves upon prior network-based methods by leveraging node embeddings learned via \DW
to prioritize \AD-associated paths.
Moreover, the use of learned embeddings allows \method to more effectively capture a network's rich topology
than previous approaches, such as \baseline.
In a direct comparison, we find that 20\% more of \method's highly prioritized drug candidates (top-50)
have published anti-AD evidence compared to \baseline's highly prioritized drug candidates.
In addition to evidence in literature,
we leverage insurance claims data to produce Cox proportional-hazard models.
Among all the drugs we evaluate, these models identified \bbbx \ace as having the lowest risk of \AD.
Similarly, etodolac was found to have the lowest risk of \AD among the four \cox we evaluated (Figure~\ref{fig:coxinh}),
indicating that this drug in particular may have protective effect despite the class as a whole
not exhibiting a significantly reduced risk of \AD compared to our negative control (Figure~\ref{fig:cox}).
Additionally, \method highly prioritizes drugs that target the cholinergic system.
Each of the approved \AD drugs in our dataset that are also \ache are prioritized in the top-50 by \method.
\method also highly prioritizes nicotine, an \acr.
This prioritization is supported by both literature and our Cox models,
which suggest nicotine is associated with reduced risk of \AD.
Altogether, the results presented in this work highlight
etodolac, nicotine, and \ace as viable candidates for repurposing to treat \AD
and, as such, deserve further examination in future studies.

Despite its promising results, \method exhibits a few limitations.
The PPI network we construct is a simplification of molecular pathways.
Like many other network-based approaches,
\method does not consider loops nor the directionality of PPI
as these can be difficult for models to learn.
In our context, this means that highly ranked candidates are only likely
to be in close proximity to \AD-related genes.
To improve drug prioritization, models must be capable of identifying drugs
that are both upstream of and in close proximity to these \AD-related genes.
In future studies, we will leverage directed interactions either by hard coding them or learning them.
One way directionality might be learned is through the use of {multi-omics} data.
Examining how changes to genomic and epigenomic profiles affect gene expression
could facilitate learning where genes are in pathways.
Furthermore, by leveraging {multi-omics} data, we may be able to provide more personalized drug recommendations.

\section{Acknowledgments}
\label{sec:acknowledgments}
This project was made possible, in part, by support from the National Institute of Aging grant no. 5R01AG071470. Any opinions, findings and conclusions or recommendations expressed in this paper are those of the authors and do not necessarily reflect the views of the funding agency.

\bibliographystyle{ws-procs11x85}
\bibliography{ws-pro-sample}

\end{document}